\documentclass[pdflatex,aps,twocolumn,superscriptaddress,prl,fleqn,showpacs,nofootinbib,preprintnumbers]{revtex4}
\usepackage{amssymb,amsmath,epsfig}

\DeclareMathOperator{\tr}{Tr}

\newcommand{\xB}{x_{\scriptscriptstyle B}}
\newcommand{\sT}{{\scriptscriptstyle T}}
\newcommand{\bm}[1]{\mbox{\boldmath $#1$}}

\begin{document}

\title{Direct Probes of Linearly Polarized Gluons inside Unpolarized
Hadrons}

\author{Dani\"el Boer}
\email{D.Boer@rug.nl}
\affiliation{Theory Group, KVI, University of Groningen,
Zernikelaan 25, NL-9747 AA Groningen, The Netherlands}

\author{Stanley J.~Brodsky}
\email{sjbth@slac.stanford.edu}
\affiliation{SLAC National Accelerator Laboratory, Stanford University,
Stanford, California 94309, USA}
\affiliation{CP$^3$-Origins, Southern Denmark University, Odense, Denmark}

\author{Piet J. Mulders}
\email{mulders@few.vu.nl}
\affiliation{Department of Physics and Astronomy, Vrije Universiteit
  Amsterdam, NL-1081 HV Amsterdam, The Netherlands}

\author{Cristian Pisano}
\email{cristian.pisano@ca.infn.it}
\affiliation{Dipartimento di Fisica, Universit\`a di Cagliari, and INFN, Sezione di Cagliari, I-09042 Monserrato (CA), Italy}

\begin{abstract}
We show that linearly polarized gluons inside
unpolarized hadrons can be directly probed in jet or heavy quark pair
production in electron-hadron  collisions. We discuss the simplest $\cos 2\phi$ asymmetries 
and estimate their maximal value, concluding that measurements of the unknown linearly polarized 
gluon distribution in the proton should be feasible in future EIC or LHeC experiments.
Analogous asymmetries in hadron-hadron collisions suffer from factorization breaking contributions and 
would allow to quantify the importance of initial and final state interactions. 
\end{abstract}

\date{\today}

\preprint{SLAC-PUB-14294}

\maketitle

Although quarks and gluons are confined within hadrons, tests of their fundamental
properties are possible through scattering processes. It has become clear that quarks are in 
general spin-polarized, even within unpolarized hadrons, 
with polarization directions and magnitudes that depend on their transverse momentum and flavor.
This nontrivial feature of hadron structure shows itself
through specific angular asymmetries in scattering processes~\cite{Boer:1997nt,Boer:1999mm,Boer:2002ju}
that have been studied in 
a number of experiments~\cite{Falciano:1986wk,Guanziroli:1987rp,Conway:1989fs,Zhu:2006gx}. 
Quark spin-polarization in unpolarized hadrons
is also supported by first-principle lattice QCD calculations \cite{Brommel:2007xd}.
What has received much less attention is that gluons can exhibit a
similar property, i.e.\ they can be linearly polarized
inside an unpolarized hadron. In this letter we propose measurements which
are directly sensitive to this unexplored gluon distribution.
Its accurate measurement would allow one to take advantage of 
polarized scattering at colliders without polarized beams.  

Thus far experimental and theoretical investigations of gluons
inside hadrons have focussed on their momentum and helicity distributions.
The gluon density $g(x)$ describing the distribution of
unpolarized gluons with a collinear momentum fraction $x$ in an unpolarized hadron,
integrated over transverse momentum $\bm p_\sT$, has been
extracted with considerable precision from measurements of high energy
electron-proton collisions at HERA (DESY).  This distribution enters the
structure function $F_L$ in inclusive deep inelastic scattering (DIS) at
order $\alpha_s$, and it drives the evolution
of sea quark distributions at small values of $x$.
The unintegrated gluon distribution $g(x, \bm p^2_\sT) $ 
enters less inclusive reactions where the transverse momentum of the gluons is taken into
account, such as semi-inclusive deep inelastic
scattering or dijet production in hadronic collisions. In these cases
the gluons are not necessarily unpolarized, even if the parent hadron
itself is unpolarized. In fact, because of their spin-orbit
couplings, the gluons can obtain a linear polarization.
This gives access to a different polarization mode compared to 
the helicity distribution $\Delta g(x)$, which is the distribution of 
circularly polarized gluons inside {\it polarized} nucleons. 

Information on linearly polarized gluons in a hadron is formally encoded in the
hadron matrix element of a correlator of the gluon field strengths $F^{\mu \nu}(0)$ and
$F^{\nu \sigma}(\lambda)$ evaluated at fixed light-front time
$\lambda^+ =\lambda{\cdot}n=0$, where $n$ is a lightlike vector
conjugate to the parent hadron's four-momentum $P$.
Specifically, the gluon content of an unpolarized hadron at
leading twist (omitting gauge links) 
for a gluon momentum 
$p = x\,P + p_\sT + p^- n$
is described by the correlator \cite{Mulders:2000sh}
\begin{eqnarray}
\label{GluonCorr}
\Phi_g^{\mu\nu}(x,\bm p_\sT )
& = &  \frac{n_\rho\,n_\sigma}{(p{\cdot}n)^2}
{\int}\frac{d(\lambda{\cdot}P)\,d^2\lambda_\sT}{(2\pi)^3}\
e^{ip\cdot\lambda}\, \nonumber
\\
& & \hspace{-0.4 cm} \times
\langle P|\,\tr\big[\,F^{\mu\rho}(0)\,
F^{\nu\sigma}(\lambda)\,\big]
\,|P \rangle\,\big\rfloor_{\text{LF}} \nonumber \\
&& \hspace{-2 cm} =
\frac{-1}{2x}\,\bigg \{g_\sT^{\mu\nu}\,f_1^g
-\bigg(\frac{p_\sT^\mu p_\sT^\nu}{M^2}\,
{+}\,g_\sT^{\mu\nu}\frac{\bm p_\sT^2}{2M^2}\bigg)
\;h_1^{\perp\,g} \bigg \}\, ,
\end{eqnarray}
with $p_{\sT}^2 = -\bm p_{\sT}^2$, $g^{\mu\nu}_{\sT} = g^{\mu\nu}
- P^{\mu}n^{\nu}/P{\cdot}n-n^{\mu}P^{\nu}/P{\cdot}n$.
This defines the
transverse momentum dependent distribution functions (TMDs)
$f_1^g(x,\bm{p}_\sT^2)$ representing the
unpolarized gluon distribution $g(x, \bm p^2_\sT)$,
at fixed light-front time, whereas
$h_1^{\perp\,g}(x,\bm{p}_\sT^2)$ is the 
distribution of linearly polarized gluons in an unpolarized hadron.  It is
named $h_1^{\perp\, g}$, because of its
resemblance to the transversely polarized quark distribution inside an
unpolarized hadron $h_1^{\perp \, q}$ (also frequently referred to as
Boer-Mulders function) \cite{Boer:1997nt}. There are
notable differences though: the $T$-odd distribution $h_1^{\perp\,q}$ for
quarks is a chiral-odd distribution (chirality-flip) and it is also
odd in $p_\sT$ (it enters as a rank 1 tensor).  It is zero in the
absence of initial or final state interactions (ISI/FSI)
\cite{Collins:2002kn,Brodsky:2002cx,Belitsky:2002sm}. The $T$-even
distribution $h_1^{\perp\,g}$ for gluons describes a $\Delta L=2$ helicity-flip
distribution, through a second rank tensor in the relative transverse momentum
$p_\sT$ ($p_\sT$-even). Since an imaginary phase is not required 
for $T$-even functions, it can in principle be nonzero in the absence of ISI or FSI. 
Nevertheless, as any TMD, $h_1^{\perp\, g}$ can receive contributions from ISI 
or FSI, leading to process-dependent gauge links in Eq.\ (\ref{GluonCorr}).
Therefore $h_1^{\perp\, g}$ can be non-universal
and its extraction can be hampered in nonfactorizing cases.

Thus far no experimental studies of the function
$h_1^{\perp\,  g}$ have been performed.  It has
been pointed out \cite{Boer:2009nc} that it contributes to the
so-called dijet imbalance in hadronic collisions, which is commonly
used to extract the average partonic intrinsic transverse momentum.
Here it enters the observable as a
convolution of two $h_1^{\perp\, g}$ functions, similarly to the
double Boer-Mulders effect which leads to a
large $\sin^2\theta \cos 2 \phi$ term 
and the leading-twist violation of the Lam-Tung relation in
Drell-Yan lepton pair production~\cite{Boer:1999mm,Boer:2002ju}.
Although in principle it is possible to isolate the contribution from the
$h_1^{\perp\, g}$ functions by appropriate
weighting of the planar angular distribution,
that is likely too difficult to do in practice. Moreover, it is unclear 
whether this weighted observable factorizes to begin with, because of
factorization breaking effects such as discussed in Ref.~\cite{Rogers:2010dm}.

Given its unique nature, it would be very interesting to obtain an
extraction of $h_1^{\perp\, g}$ in a simple and theoretically safe 
manner. This turns out to be possible, since unlike
$h_1^{\perp\, q}$, it does not need to appear in pairs. In this letter we will
discuss several new ways to probe the linear gluon polarization using
observables that involve only a single $h_1^{\perp\, g}$. 
The processes of interest, semi-inclusive DIS to two heavy quarks or to two jets,
allow for TMD factorization and hence a safe extraction. Analogous processes
in proton-proton collisions run into the problem 
of factorization breaking. A difference between the extractions will 
allow to quantify the importance of ISI/FSI.

We first consider the electroproduction of heavy quarks,
$e (\ell){+}h(P)\to e(\ell^\prime) {+} Q(K_1) {+} \bar{Q}(K_2){+}X$,
where the four-momenta of the particles are given within brackets, and
the quark-antiquark pair in the final state is almost back-to-back in the
plane perpendicular to the direction of the exchanged photon and hadron.
The calculation proceeds along the lines explained
in Refs.\ \cite{Boer:2007nd,Boer:2009nc}.
We obtain for the cross section
integrated over the angular distribution of the back-scattered
electron $e(\ell^\prime)$:
\begin{eqnarray}
\frac{d\sigma}
{dy_1\,dy_2\,dy\,d\xB\,d^2\bm{q}_{\sT} d^2\bm{K}_{\perp}} =  
\delta(1 - z_1 - z_2) \qquad \qquad \,\,&& \nonumber \\
\!\!\!\times \frac{\alpha^2\alpha_s}{\pi  s
  M_\perp^2}\, \frac{(1+y \xB)}{ y^5 \xB}\, \bigg[ A +\frac{\bm q_\sT^2}{M^2}\,
  B \, \cos 2 (\phi_\sT-\phi_\perp) \bigg].\,\,\, &&
\label{eq:cso}
\end{eqnarray}
This expression involves the standard DIS variables:
$Q^2 = -q^2$, where $q$ is the momentum of the virtual photon,
$\xB = Q^2/2P\cdot q, y = P\cdot q/P\cdot \ell$ and
$s = (\ell + P)^2 = 2\,\ell\cdot P = 2\,P\cdot q/y = Q^2/\xB y$.
Furthermore, we have for the jet momenta $K_{i\perp}^2 = -\bm
K_{i\perp}^2$ and introduced the rapidities $y_i$ for the 
heavy quark (HQ) or jet momenta
(along photon-target direction). We denote the heavy (anti)quark
mass with $M_Q$. For the partonic subprocess we have
$p+q=K_1+K_2$, implying $z_1+z_2 = 1$, where $z_i=P\cdot K_i/P\cdot q$.
We introduced the  sum and difference of the transverse HQ or jet  
momenta, $K_\perp = (K_{1\perp} - K_{2\perp})/2$ and
$q_\sT = K_{1\perp} + K_{2\perp}$ with
$\vert q_\sT\vert \ll \vert K_\perp\vert$. In that situation,
we can use the approximate transverse HQ or jet momenta
$K_{1\perp} \approx K_{\perp}$ and $K_{2\perp} \approx -K_{\perp}$
denoting $M_{i\perp}^2 \approx M_\perp^2 = M_Q^2 + \bm K_\perp^2$.
The azimuthal angles of $\bm{q}_\sT$ and  $\bm{K}_\perp$ are denoted
by $\phi_\sT$ and $\phi_\perp$, respectively.
The functions $A$ and $B$
in general depend on $\xB, y, z (\equiv z_2), Q^2/M_{\perp}^2,
M_Q^2/M_{\perp}^2$, and  $\bm{q}_\sT^2$.

The explicit expression for the angular independent part $A$
involves only $f_1^g$. We will focus here on the coefficient $B$ of the
$\cos 2 ( \phi_\sT-\phi_\perp)$ angular distribution and we obtain
\begin{equation}
B^{e h\to e {Q} \bar{Q}  X}=
\sum_Q e^2_Q\,  h_1^{\perp \,g} (x, \bm{q}_{\sT}^2)
{\cal B}^{e  g\to e Q \bar Q}\,  ,
\label{eq:BQQb}
\end{equation}
with
\begin{align}
& {\cal B}^{e  g\to e Q\bar Q} = \frac{1}{2}\, \frac{z (1-z)}{D^3}\, \left
  (1-\frac{M_Q^2}{M_\perp^2} \right ) a(y)  \nonumber\\
& \times
\left [ \big (2\, z (1-z) \, b(y) - 1\big )
  \frac{Q^2}{M_\perp^2} +2 \,\frac{M_Q^2}{M_\perp^2}\right]~,
\label{eq:BQQb2}
\end{align}
$D \equiv D \left (z,Q^2/M_\perp^2 \right ) = 1 + z (1-z) Q^2/M_\perp^2$,
$a(y) = 2 -y (2-y)$, 
$b(y) =  (6 -y (6-y))/a(y).$

One observes that the magnitude $B$ of the $\cos 2 \phi$ asymmetry,
where $\phi=\phi_\sT-\phi_\perp$, is determined by $h_1^{\perp\, g}$
and that if $Q^2$ and/or $M_Q^2$ are of the same order as
$K_\perp^2$, the coefficient $B$ is not power suppressed.
Since $h_1^{\perp\, g}$ is completely unknown, we estimate the maximum
asymmetry that is allowed by the bound:
\begin{equation}
|h_1^{\perp\, g (2)}(x)| \leq \frac{\langle p_\sT^2\rangle}{2M^2} \
 f_1^g(x)\, ,
\label{bound}
\end{equation}
that we derived from the spin density matrix given in
\cite{Mulders:2000sh} in the way presented in
Ref.\ \cite{Bacchetta:1999kz}.
The superscript $(2)$ denotes the $n=2$ transverse moment.
Transverse moments of TMDs are defined as:
$f^{(n)}(x) \equiv
\int  d^2 \bm{p}_\sT\,
\left(\bm{p}_\sT^2/2 M^2\right)^n
\,f(x, \bm{p}_\sT^2)$  
(a suitably chosen regularization is understood, e.g.\ as discussed 
in appendix B of \cite{Bacchetta:2008xw}).
If we select 
$Q^2 = M_Q^2=K_\perp^2/4$, 
$y_1=y_2$, the asymmetry ratio
\begin{equation}
\left| \frac{\int d^2 \bm{q}_\sT
\, \bm{q}_\sT^2
\, \cos 2 (\phi_\sT-\phi_\perp) \, d\sigma}{\int d^2 \bm{q}_\sT
\, \bm{q}_\sT^2
\, d\sigma}\right| =  \frac{\int d\bm{q}_\sT^2 \ \bm{q}_\sT^4\, | B |}{
2 M^2 \int d\bm{q}_\sT^2 \ \bm{q}_\sT^2 \, A} \, ,
\end{equation}
is maximally around 13\%, which we view as encouraging.

If one
keeps the lepton plane angle $\phi_\ell$, there
are other azimuthal dependences such as a $\cos 2(\phi_\ell - \phi_\sT)$, 
but its bound 
is at least 6 times smaller than on $\cos 2(\phi_\sT - \phi_\perp)$.

The cross section for the process $e\, h\to e^\prime\,{\rm jet}\, {\rm jet}\, X$
can be calculated in a similar way. The corresponding expressions
can be obtained from Eqs.\ (\ref{eq:BQQb}) and (\ref{eq:BQQb2})
with $M_Q=0$. One can
then also replace the rapidities of the outgoing particles, $y_i$, with
the pseudo-rapidities $\eta_i {=}\,{-}\ln\left[\tan(\frac{1}{2}\theta_i)\right]$,
$\theta_i$ being the polar angles of the final partons in the
virtual photon-hadron cms frame. 
Note that $A$ now also receives a contribution 
from $\gamma^* q \to g q$, 
leading to somewhat smaller asymmetries. 

Since the observables involve final-state heavy quarks or jets, they require
high energy colliders, such as a future Electron-Ion Collider (EIC) or 
the Large Hadron electron Collider (LHeC) proposed at CERN.
It is essential that 
the individual transverse momenta $K_{i\perp}$ are reconstructed with an accuracy $\delta K_\perp$
better than the magnitude of the sum of the transverse momenta 
$K_{1\perp}+K_{2\perp}=q_{\scriptscriptstyle T}$. Thus one has to satisfy 
$\delta K_\perp \ll \vert q_{\scriptscriptstyle T}\vert \ll \vert K_\perp\vert$.  

An analogous asymmetry
arises in QED, in the `tridents' processes $\ell e(p) \to \ell \mu^+ \mu^- e^\prime (p^\prime\,{\rm or}\, X)$ or
$\mu^- Z \to \mu^- \ell \bar{\ell} Z$ \cite{Bjorken:1966kh,Brodsky:1966vh,Tannenbaum:1968zz,Gluck:2002cm}.
This could be described by the distribution of linearly polarized photons inside a lepton, proton, or atom.
QCD adds the twist that for gluons inside a hadron, ISI or FSI can considerably modify the
result depending on the process, for example, in HQ production in
hadronic collisions: $p\, p \to Q\, \bar{Q}\, X$, 
which can be studied at BNL's Relativistic Heavy Ion Collider (RHIC)
and CERN's LHC, and $p\, \bar{p} \to Q\, \bar{Q}\, X$  at Fermilab's Tevatron. 
Since the description involves two TMDs, breaking of TMD factorization becomes 
a relevant issue, cf.\ \cite{Rogers:2010dm} and references therein.
The cross section for the process
$h_1(P_1){+}h_2(P_2)\, {\rightarrow}\,Q(K_1){+} \bar{Q}(K_2){+}X$
can be written in a way similar to the hadroproduction of two jets discussed in
Ref.\ \cite{Boer:2009nc}, in the following form
\begin{align}
& \frac{d\sigma}
{dy_1 dy_2 d^2\bm{K}_{1\perp} d^2\bm{K}_{2\perp}}  =
\frac{\alpha_s^2}{s {M}_\perp^2} \nonumber \\
& \quad \times \Big[ A(\bm{q}_\sT^2) + B(\bm{q}_\sT^2) \bm q_\sT^2 \cos 2 (\phi_\sT-\phi_\perp)
\nonumber\\
& \qquad \quad + C(\bm{q}_\sT^2) \bm{q}_\sT^4 \cos 4 (\phi_\sT-\phi_\perp)\Big]\, .
\label{eq:csoQQb}
\end{align}
Besides $\bm q_\sT^2$, the terms $A$, $B$ and $C$ 
will depend on other, often not explicitly indicated, variables as
$z$, $M_Q^2/M_\perp^2$ and momentum fractions $x_1$, $x_2$ obtained from 
$x_{1/2} = \left(\,M_{1 \perp}\,e^{\pm y_1}\,
{+} M_{2 \perp}\,e^{\pm y_2}\,\right )/ \sqrt{s}$ .

\begin{figure}[b]
\centering
\psfig{file=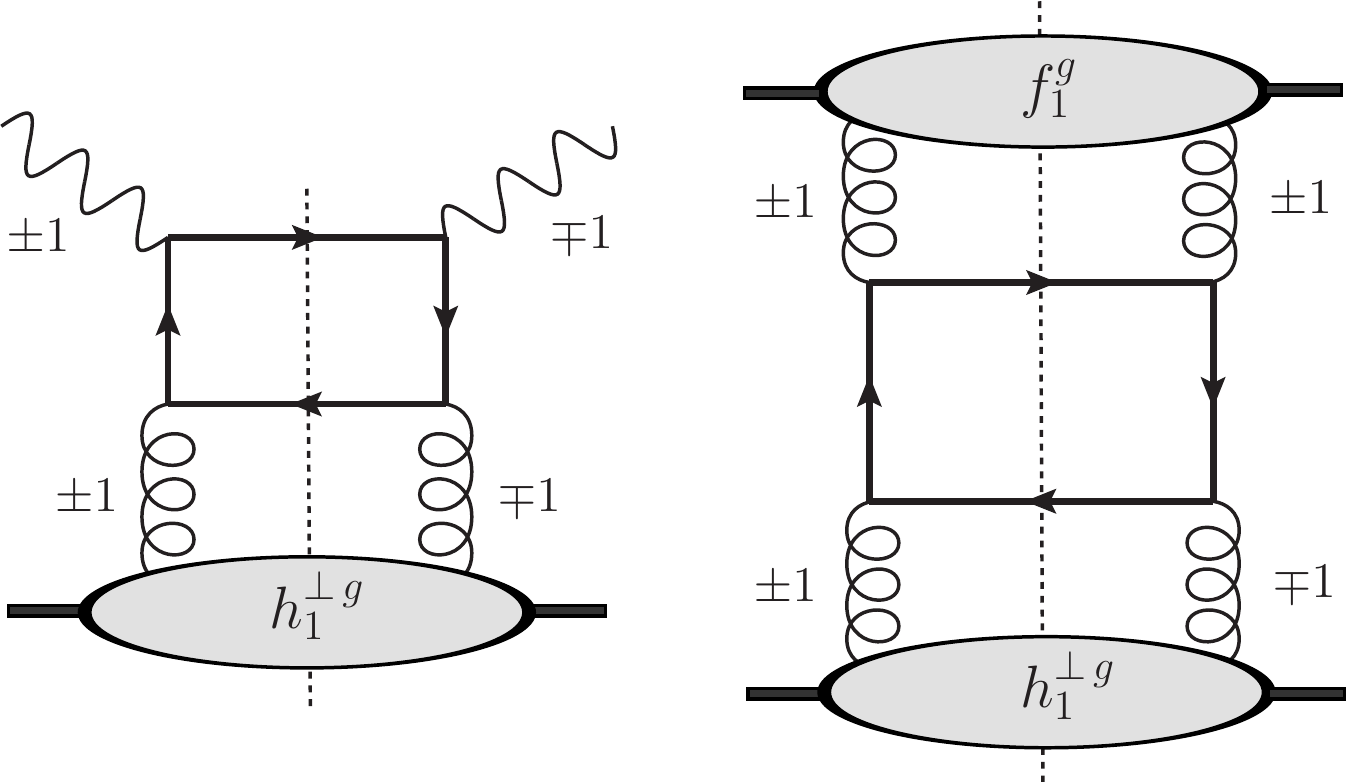, width=0.45\textwidth}
\caption{\it Examples of subprocesses contributing
to the $\cos 2 \phi$ asymmetries in $e\, p \to e^\prime \,Q\, \bar{Q}\, X$
and $p\, p \to Q\, \bar{Q}\, X$, respectively. As the helicities of the
photons and gluons indicate, the latter process requires helicity
flip in quark propagators resulting in an $M_Q^2/M_\perp^2$ factor.}
\label{diagrams}
\end{figure}

In the most naive partonic description the terms A, B, and C contain 
convolutions of TMDs. Schematically,
\begin{eqnarray}
A: && f_1^q \otimes f_1^{\bar{q}},\ f_1^g \otimes f_1^g\,,\nonumber\\
B: && h_1^{\perp \, q} \otimes h_1^{\perp \, {\bar q}},\
\frac{M_Q^2}{M_\perp^2} f_1^{g} \otimes h_1^{\perp \,
  g}\,, \nonumber\\ 
C: && h_1^{\perp \, g} \otimes h_1^{\perp \, g}~. \nonumber 
\end{eqnarray}
Terms with higher powers in $M_Q^2/M_\perp^2$ are left out. In Fig.~\ref{diagrams} 
the origin of the factor $M_Q^2/M_\perp^2$ in
the contribution of $h_1^{\perp\, g}$ to $B$ is explained.

The factorized description in terms of TMDs is problematic though. 
In Ref.~\cite{Rogers:2010dm} it was pointed out that for hadron or jet pair production in 
hadron-hadron scattering TMD factorization fails. The ISI/FSI will not allow a separation of 
gauge links into the matrix elements of the various TMDs. Only in 
specific simple cases, such as the single Sivers effect, one can find weighted
expressions that do allow a factorized result, but with in general different factors for 
different diagrams in the partonic subprocess~\cite{Bomhof:2006dp,Bomhof:2007xt}.
Even if this applies to the present case for $A$ and $B$ as well, 
actually two different functions $h_{1}^{\perp g (2)}(x)$ (and $f_1^{g (1)}(x)$) will appear,
corresponding to gluon operators with the color structures 
$f_{abe}\,f_{cde}$ and $d_{abe}\,d_{cde}$, 
respectively~\cite{Bomhof:2007xt,Dominguez:2010xd}.
This is similar to what happens for single transverse spin asymmetries ($A_N$) in heavy
quark production processes 
\cite{Yuan:2008vn,Kang:2008qh,Yuan:2008it,Kang:2008ih,Anselmino:2004nk}. 
Because there too two different ($f$ and $d$ type) gluon correlators arise, 
the single-spin asymmetries in $D$ and $\bar{D}$ meson production
are found to be different. However, in the unpolarized scattering case considered in this
letter the situation is simpler, since only one operator contributes or 
dominates. In the $\gamma^* g \to \,Q\,\bar Q$ subprocess only the matrix element with
the $f\,f$-structure appears, while in the $g\,g\to Q\,\bar Q$ subprocess relevant for 
hadron-hadron collisions the $d\,d$-structure dominates (the $ff$-contribution is suppressed
by $1/N^2$). A side remark on $p_\sT$ broadening \cite{Luo:1993ui,Guo:1998rd,Liang:2008vz}: 
because of the two different four-gluon operators for $f_1^{g (1)}(x)$ we expect the broadening 
$\Delta p_\sT^2$ in SIDIS,
$(\Delta p_\sT^2)_{\rm DIS} \equiv \langle p_\sT^2 \rangle_{eA} - \langle p_\sT^2 \rangle_{ep}$,
to be different from the one in hadron-hadron collisions,
$(\Delta p_\sT^2)_{\rm hh} \equiv \langle p_\sT^2 \rangle_{pA} - \langle p_\sT^2 \rangle_{pp}$.

In case weighting does allow for factorized expressions, we present here the 
relevant expressions for $B={\cal B}^{q \bar q \to Q \bar{Q}} + (M_Q^2/M_\perp^2)\, {\cal B}^{g g \to Q \bar{Q}}$, where
\begin{eqnarray}
{\cal B}^{q  \bar q \to Q\bar Q } & = & \frac{N^2-1}{N^2}\,{z}^2
(1-z)^2  \left (1 -\frac{M_Q^2}{M_\perp^2} \right )\nonumber \\
& & \hspace{-1 cm}\times \bigg [ {\cal{H}}^{q \bar{q}}(x_1, x_2,\bm{q}_{\sT}^2) + {\cal{H}}^{\bar q q}(x_1, x_2,\bm{q}_{\sT}^2) \bigg ]\, ,\nonumber \\
{\cal B}^{g g  \to Q\bar Q } &=& \frac{N}{N^2-1}\,{\cal B}_1\, {\cal{H}}^{g g }(x_1, x_2,\bm{q}_{\sT}^2)\, ,
\end{eqnarray}
with $N$ being the number of quark colors and
\begin{equation}
{\cal B}_1 ={z} (1-z)
\left ({z}^2 +(1-z)^2 -\frac{1}{N^2}\right )
  \left (1 -\frac{M_Q^2}{M_\perp^2} \right )~.
\end{equation}
The weighted integrals which appear in
the $\bm{q}_\sT^2/M^2$-weighted cross section
(cf.\ \cite{Boer:2009nc}), for $M_1=M_2=M$ are:
\begin{align}
& \pi\int d\bm{q}_\sT^2\ \left(\frac{\bm{q}_\sT^2}{M^2}\right)
\,\bm{q}_\sT^2\, {\cal{H}}^{q \bar q }(x_1, x_2, \bm{q}_\sT^2)
 =  \nonumber\\
& \qquad
8\sum_{{\rm flavors}}\,h_1^{\perp q(1)}(x_1)\,h_1^{\perp \bar
  q(1)}(x_2)\, ,
 \label{h1perpqsquared}
\end{align}
already discussed in \cite{Boer:2009nc}, and
\begin{align}
&\pi\int d\bm{q}_\sT^2
\ \left(\frac{\bm{q}_\sT^2}{M^2}\right)
\,\bm{q}_\sT^2\, {\cal{H}}^{g g  }(x_1, x_2, \bm{q}_\sT^2)
=\nonumber \\
& \quad 4 \left (h_1^{\perp g (2)}(x_1)\,f_1^{ g }(x_2)
+ f_1^{ g }(x_1)\,h_1^{\perp g (2)}(x_2) \right )\, .
\label{weightedconvol}
\end{align}
Whether $g g \to Q \bar{Q}$ is more important than 
$q  \bar q \to Q\bar Q$, depends
strongly on the values of $x_i$ and $M_Q^2/M_\perp^2$,
and on whether one deals with $p\,p$ or $p\,\bar{p}$.
In $p \,\bar{p}$ collisions and for $K_\perp^2$ not too large compared to $M_Q^2$, the
contribution from $h_1^{\perp\, g}$ is expected to be the dominant one.
The importance of the contribution from
$h_1^{\perp \, q}$ can be assessed through a
comparison to photon-jet production \cite{Boer:2007nd}.

In summary, measurements of the azimuthal asymmetry of jet or heavy quark pair production in 
$e \,p$ and in $p\, p$ or $p \,\bar{p}$ collisions (and possibly also in diphoton or even Higgs 
production \cite{Nadolsky:2007ba,Catani:2010pd})
can directly probe the distribution of linearly polarized gluons inside unpolarized hadrons.
From a theoretical viewpoint 
this asymmetry in the $e\, p$ process is among the simplest TMD observables. 
Breaking of TMD factorization is expected in the $p\,p$ case and may lead to
uncontrolled corrections. A comparison between extractions from these two types of processes 
would therefore be very interesting.  
The relative simplicity of the proposed measurements (polarized beams are not
required) suggests a promising prospect for the extraction of this 
gluon distribution in the future and for the study of its potential process dependence.

\begin{acknowledgments}
We thank John Collins, Markus Diehl, Francesco Murgia, Ted Rogers, and Jianwei Qiu for useful
discussions.
C.P.~is supported by Regione Autonoma della Sardegna (RAS) through a research
grant under the PO Sardegna FSE 2007-2013, L.R. 7/2007, ``Promozione della
ricerca scientifica e dell'innovazione tecnologica in Sardegna".
This research is part of the FP7 EU-programme Hadron Physics (No.\ 227431).
\end{acknowledgments}

\end{document}